\documentstyle[aps,preprint,eqsecnum,epsf]{revtex}
\begin{document}
\draft
\title{Stability of Uniform Shear Flow}
\author{Jos\'{e} M. Montanero}
\address{Departamento de Electr\'onica e Ingenier\'{\i}a Electromec\'anica,
Universidad de Extremadura, \\
E-06071 Badajoz, Spain}
\author{Andr\'{e}s Santos}
\address{Departamento de F\'{\i}sica, 
Universidad de Extremadura, \\
E-06071 Badajoz, Spain}
\author{Mirim Lee}
\address{TCSUH, Department of Physics, University of Houston,\\
Houston, TX 77204}
\author{James W. Dufty}
\address{Department of Physics, University of Florida \\
Gainesville, FL 32611}
\author{J. F. Lutsko}
\address{ESADG, Department of Chemical Engineering, Katholiek University of 
Leuvan,\\B-3001, Heverlee, Belgium}
\date{\today }
\maketitle

\begin{abstract}
The stability of idealized shear flow at long wavelengths is studied in
detail. A hydrodynamic analysis at the level of the Navier-Stokes equation
for small shear rates is given to identify the origin and universality of an
instability at any finite shear rate for sufficiently long wavelength
perturbations. The analysis is extended to larger shear rates using a low
density model kinetic equation. Direct Monte Carlo simulation of this
equation is compared with a hydrodynamic description including non Newtonian
rheological effects. The hydrodynamic description of the instability is in
good agreement with the direct Monte Carlo simulation
for $t<50t_0$, where $t_0$ is the
mean free time. Longer time simulations up to $2000t_0$ are used to identify
the asymptotic state as a spatially non-uniform quasi-stationary state.
Finally, preliminary results from molecular dynamics simulation showing the
instability are presented and discussed.
\end{abstract}


\section{Introduction}

\label{sec:intro}

Uniform shear flow is a prototype non-equilibrium state admitting detailed
study at both the macroscopic and microscopic levels via theory and computer
simulation. This is an idealized version of shear flow between parallel
plates in which the velocity profile is exactly linear in a coordinate
orthogonal to the flow direction (as in Couette flow) and with a spatially
uniform temperature and pressure (in contrast to Couette flow). There is a
single scalar control parameter, the shear rate $a$, which measures how far
the system is driven from equilibrium. This flow is generated by periodic
boundary conditions in the local Lagrangian frame (Lees-Edwards boundary
conditions) that can be implemented at the levels of hydrodynamics, kinetic
theory, and Newtonian mechanics \cite{edwards,dufty1,dufty2}. Although these
boundary conditions are non-local and therefore not reproducible in real
experiments, they are ideally suited for computer simulation of this special
nonequilibrium state and for more penetrating theoretical analysis. In this
way, a quantitative study of rheological properties usually associated with
complex molecular systems has been performed for simple atomic fluids \cite
{shear}. The most complete studies have been via molecular dynamics
simulation of Newtonian dynamics at high densities and, more recently, by
Monte Carlo simulation of the Boltzmann equation at low densities \cite
{gomez,MSG96}. Molecular dynamics simulations have revealed a transition
from fluid symmetry to an ordered state at sufficiently high shear rates 
\cite{erpenbeck}, which has been attributed to a short wavelength
hydrodynamic instability \cite{lutsko}. The objective here is to show that
uniform shear flow also is unstable at sufficiently long wavelengths, for
any finite value of the shear rate \cite{mirim:2,mirim:3}. This instability
has not been seen in earlier computer simulations due to the small system
sizes considered, with consequent restrictions to shorter wavelengths. The
instability is identified theoretically from a hydrodynamic analysis both
near and far from equilibrium. This analysis is confirmed quantitatively at
short times by Monte Carlo simulations of an associated low density kinetic
equation. The asymptotic evolution of this instability also is explored via
Monte Carlo simulation showing transition to a non-steady, spatially
inhomogeneous state superimposed upon the uniform shear flow. Finally, the
instability is demonstrated by new molecular dynamics simulations on larger
systems to allow consideration of long wavelengths. This instability does
not invalidate or compromise in any way previous studies of uniform shear
flow, since we also establish that the flow is stable at sufficiently short
wavelengths.

The boundary conditions generate viscous heating so that uniform shear flow
is not stationary. This viscous heating can be controlled by the
introduction of a non-conservative external force that acts as a uniform
thermostat. The simplest choice is a Stokes law drag force on each particle
proportional to its velocity relative to the local macroscopic flow. The
proportionality ``constant'' then is adjusted to compensate for the heating.
There are several possibilities in the detailed implementation of the
thermostat, leading to the same properties for the stationary state but
different hydrodynamics for small perturbations from that state. The theory
and Monte Carlo simulations are carried out for both global and local
thermostats. The role of the thermostat is studied in the Appendix  where 
it is shown that the qualitative features of the instability are not 
sensitive to the choice of thermostat.

In the next section, the usual Navier-Stokes hydrodynamic equations are
considered. These equations are restricted to small gradients of the
hydrodynamic fields relative to equilibrium, and consequently the shear rate
must be small in this analysis. Otherwise, the density and interatomic force
law can be considered arbitrary within the fluid phase. The stationary
solution for uniform shear flow is identified, and the linear hydrodynamic
equations for small perturbations of this solution are studied. The five
hydrodynamic modes are determined in detail for the special case of spatial
perturbations orthogonal to the flow. A critical wavevector, $k_c(a)$, is
determined such that for wavevectors $k<k_c(a)$ the perturbations grow as a
function of time. The critical wavevector vanishes as the shear rate $a$
goes to zero, but for any finite value of the shear rate there are
sufficiently small wavevectors (long wavelengths) such that the perturbation
is unstable. 

These hydrodynamic predictions are tested by comparison to Monte Carlo
simulations at the more fundamental kinetic theory level. A model kinetic
equation has been analyzed for states near uniform shear flow, without
restriction on the shear rate \cite{mirim:3}. The hydrodynamic equations for
small deviations from uniform shear flow determine the critical wavevector, $%
k_c(a)$, for values of the shear rate $a$ beyond the limitations of the
Navier-Stokes equations where efficient Monte Carlo simulations are
possible. The theoretical prediction of the growth of initial perturbations
is compared with a direct \ Monte Carlo simulation of a solution to the
kinetic equation. The results confirm both the hydrodynamic analysis and the
prediction of an instability for times up to about $50t_0$, where $t_0$ is
the mean free time, after which the initial growth has exceeded the
limitations of the linear stability analysis. The simulation results are
continued up to $2000t_0$ to explore the ultimate stabilization by
non-linear effects. The asymptotic state for the hydrodynamic fields appears
to be a system size dependent standing wave with a period of about $50t_0$.
Further details and discussion are given in section \ref{sec:monte}.

In section \ref{sec:mole} some preliminary attempts to see the instability
at high densities using molecular dynamics for the hard sphere fluid are
discussed. The system dimension in the direction of the spatial perturbation
is increased by an order of magnitude relative to previous simulations. A
long wavelength perturbation is found to grow on a very long simulation
time, with no indication of approach to the steady uniform shear flow. A
quantitative comparison with theory at the required larger densities and
shear rates is now possible using a recently developed kinetic model for the
hard sphere Enskog equation for application \cite{dufty3}, although the
details have not been worked out at this point. The results of the theory
and simulations are summarized and discussed in section \ref{sec:dis}.

\section{Navier-Stokes Analysis}

\label{sec:navier} 
On sufficiently large space and time scales the dynamics
of a fluid is well-described by hydrodynamic equations obtained from the
exact conservation laws for the average mass, energy, and momentum
densities, together with approximate constitutive equations for the
associated fluxes. For the analysis here we use the density, $n({\bf r},t)$,
temperature, $T({\bf r},t)$, and flow velocity, ${\bf U}({\bf r},t)$, as
dependent variables rather than the density, energy density, and momentum.
The general form of these conservation laws is \cite{McL}

\begin{equation}
D_tn+n\nabla \cdot {\bf U}=0,  \label{1}
\end{equation}

\begin{equation}
D_tT+(\gamma -1)\alpha ^{-1}\nabla \cdot {\bf U}+(mnC_v)^{-1}\left( \nabla
\cdot {\bf q}+P_{ij}\partial _jU_i-w\right) =0,  \label{2}
\end{equation}

\begin{equation}
D_tU_i+\rho ^{-1}\partial _ip+\rho ^{-1}\partial _jP_{ij}=0,  \label{3}
\end{equation}
where $D_t\equiv \partial_t+{\bf U}\cdot \nabla $ is the material
derivative. The parameters occurring in these equations are the mass
density, $\rho =mn$, the specific heat at constant volume, $C_v$, the
pressure, $p$, the ratio of specific heats at constant pressure and volume, $%
\gamma =C_p/C_v $, and the coefficient of expansion, $\alpha =-n^{-1}\left(
\partial n/\partial T\right) _p$. These parameters are the same functions of
the local density and temperature as in equilibrium. Finally, the
irreversible heat and momentum fluxes are denoted by ${\bf q}$ and $P_{ij}$,
respectively, and $w=w(n,T)$ is the rate at which work is done by the
external force representing the thermostat. Its detailed form will not be
required in this section.

The above equations are incomplete until constitutive equations for the
fluxes are specified in terms of the hydrodynamic fields. However, the
special solution of uniform shear flow exists independent of this choice. It
is defined by a spatially constant temperature, density, heat flux and
momentum flux, and a flow velocity whose only non-vanishing component is $%
U_{s,x}=ay$. The shear rate, $a$, provides the single control parameter
measuring the deviation from equilibrium. The boundary conditions are simple
periodic conditions in the local Lagrangian coordinate frame, ${\bf %
r^{\prime }}={\bf r}-{\bf U}_s({\bf r})t$. Substitution of these assumptions
for $n=n_s$, $T=T_s$, and ${\bf U=U}_s$ into the above conservation laws
shows that (\ref{1}) and (\ref{3}) are satisfied, while (\ref{2}) reduces to 
\begin{equation}
\partial _tT_s=-(mn_sC_{v,s})^{-1}\left( aP_{s,xy}-w(n_s,T_s)\right) .
\label{4}
\end{equation}
This expresses the temperature evolution as a competition between the
viscous heating effect, $\propto $ $aP_{s,xy}$, and the cooling by the
thermostat, $\propto w(n_s,T_s)$. A steady state is obtained by choosing the
thermostat to cancel the viscous heating, 
\begin{equation}
aP_{s,xy}=w(n_s,T_s).  \label{5}
\end{equation}
The various thermostats described in the Appendix all satisfy (\ref{5})
but differ for states away from the steady state.

Next consider the equations for small deviations of the hydrodynamic
variables from the uniform shear flow state, retaining only terms linear in
these deviations. To proceed it is necessary to specify the constitutive
equations for the heat and momentum fluxes. In this section, attention is
limited to small spatial gradients, including the shear rate, so that
Fourier's law and Newton's viscosity law apply, 
\begin{equation}
{\bf q}=-\kappa {\bf \nabla }T,  \label{6}
\end{equation}
\begin{equation}
P_{ij}=-\eta (\partial _iU_j+\partial _jU_i-\frac 23\delta _{ij}\nabla \cdot 
{\bf U})-\eta ^{\prime }\delta _{ij}\nabla \cdot {\bf U.}  \label{7}
\end{equation}
Here $\kappa (n,T)$ is the thermal conductivity, $\eta (n,T)$ is the shear
viscosity, and $\eta ^{\prime }(n,T)$ is the bulk viscosity. It follows
immediately that
\begin{equation}
{\bf q}_s=0,\hspace{0.4in}P_{s,ij}=-\eta _sa\left( \delta _{ix}\delta
_{jy}+\delta _{iy}\delta _{jx}\right) ,  \label{8}
\end{equation}
\begin{equation}
\delta {\bf q}=-\kappa _s{\bf \nabla }\delta T,  \label{8bis}
\end{equation}
\begin{equation}
\delta P_{ij}=-\eta _s(\partial _i\delta U_j+\partial _j\delta U_i-\frac 23%
\delta _{ij}\nabla \cdot \delta {\bf U})-\eta _s^{\prime }\delta _{ij}\nabla
\cdot \delta {\bf U-}a\left( \delta _{ix}\delta _{jy}+\delta _{iy}\delta
_{jx}\right) \left( \eta _{s,n}\delta n+\eta _{s,T}\delta T\right) .
\label{9}
\end{equation}
An abbreviated notation has been introduced where the subscript $s$ on a
quantity indicates it is evaluated at $n_s,T_s$. Also, $p_{s,n}\equiv \partial
p(n_s,T_s)/\partial n_s$, $p_{s,T}\equiv \partial p(n_s,T_s)/\partial T_s$, $%
\eta _{s,n}\equiv \partial \eta (n_s,T_s)/\partial n_s$, $\eta _{s,T}\equiv
\partial \eta (n_s,T_s)/\partial T_s$, etc. With these results, the closed
set of linear hydrodynamic equations for perturbations of uniform shear flow
at small shear rates are given by 
\begin{equation}
\left( \partial _t+{\bf U}_s\cdot \nabla \right) \delta n+n_s\nabla \cdot
\delta {\bf U}=0,  \label{10}
\end{equation}
\begin{eqnarray}
\left( \partial _t+{\bf U}_s\cdot \nabla \right) \delta T+(\gamma
_s-1)\alpha _s^{-1}\nabla \cdot \delta {\bf U} &+&(mn_sC_{v,s})^{-1}\left(
-\kappa _s\nabla ^2\delta T-2\eta _sa\left( \partial _x\delta U_y+\partial
_y\delta U_x\right) \right)  \nonumber \\
&=&(mn_sC_{v,s})^{-1}\left( a^2\left( \eta _{s,n}\delta n+\eta _{s,T}\delta
T\right) +\delta w\right) ,  \label{11}
\end{eqnarray}

\begin{equation}
\left( \partial _t+{\bf U}_s\cdot \nabla \right) \delta U_i+\delta
_{ix}a\delta U_y+\rho _s^{-1}\left( p_{s,n}\partial _i\delta n+p_{s,T}\partial
_i\delta T\right) +\rho _s^{-1}\partial _j\delta P_{ij}=0.  \label{12}
\end{equation}
In this section we choose a local thermostat for which $\delta w$ in the
temperature equation compensates for the excess viscous heating due to
perturbations of the temperature and density, 
\begin{equation}
\delta w=-a^2\left( \eta _{s,n}\delta n+\eta _{s,T}\delta T\right) .
\label{thermo1}
\end{equation}
This does not imply a constant local temperature, however, except for
spatially homogeneous deviations from uniform shear flow, although it does
lead to a constant average temperature or kinetic energy for the whole
system.

These differential equations have constant coefficients which suggests an
equivalent algebraic form using Fourier and Laplace transformation. Consider
first the Fourier transform. The boundary conditions are periodic in the
local Lagrangian frame given by $r_i^{\prime }\equiv r_i-U_{s,i}({\bf r})t=
\Lambda_{ij}(t)r_j$, where $\Lambda _{ij}(t)\equiv \delta _{ij}-a\delta_{ix}
\delta_{jy}t$, so it is appropriate to define the transform with respect to
the variable ${\bf r}^{\prime }$, 
\begin{equation}
\delta \widetilde{y}_\alpha ({\bf k},t)=\int d{\bf r}^{\prime }e^{i{\bf %
k\cdot r}^{\prime }}\delta y_\alpha ({\bf r},t)=\int d{\bf r}e^{i{\bf k}(t) 
{\bf \cdot r}}\delta y_\alpha ({\bf r},t),  \label{Fourier}
\end{equation}
where $\delta y_\alpha ({\bf r},t)$ denotes the set of perturbations,
considered as a function ${\bf r}^{\prime }$ in the first equality. The
periodic boundary conditions require $k_i=2n_i\pi /L_i$, where $n_i$ are
integers and $L_i$ are the linear dimensions of the system. However, since
the time derivative in the hydrodynamic equations is taken at constant ${\bf %
r}$, the representation following the second equality is useful with $%
k_i(t)\equiv k_j\Lambda_{ji}(t)$. The Fourier transformed hydrodynamic
equations are 
\begin{equation}
\partial _t\delta \widetilde{y}_\alpha +\left( A(a)-ik(t)B(a)+k^2(t)D\right)
_{\alpha \beta }\delta \widetilde{y}_\beta =0.  \label{generaleq}
\end{equation}
The three matrices $A(a)$, $B(a)$, and $D$ are 
\begin{equation}
A_{\alpha \beta }=\,a\delta _{\alpha 3}\delta _{\beta 4},  \label{matA}
\end{equation}
\begin{equation}
B_{\alpha \beta }(a)=\left( 
\begin{array}{ccccc}
0 & 0 & n_s\widehat{k}_x & n_s\widehat{k}_y & n_s\widehat{k}_z \\ 
0 & 0 & c_1\widehat{k}_x-2a\eta _sc_2\widehat{k}_y & c_1\widehat{k}_y-2a\eta
_sc_2\widehat{k}_x & c_1\widehat{k}_z \\ 
\rho _s^{-1}\left( \widehat{k}_xp_{s,n}-\widehat{k}_ya\eta _{s,n}\right) & \rho
_s^{-1}\left( \widehat{k}_xp_{s,T}-\widehat{k}_ya\eta _{s,T}\right) & 0 & 0 & 0
\\ 
\rho _s^{-1}\left( \widehat{k}_yp_{s,n}-\widehat{k}_xa\eta _{s,n}\right) & \rho
_s^{-1}\left( \widehat{k}_yp_{s,T}-\widehat{k}_xa\eta _{s,T}\right) & 0 & 0 & 0
\\ 
\rho _s^{-1}\widehat{k}_zp_{s,n} & \rho _s^{-1}\widehat{k}_zp_{s,T} & 0 & 0 & 0
\end{array}
\right) ,  \label{matB}
\end{equation}

\begin{equation}
D_{\alpha \beta }=\rho _s^{-1}\left( 
\begin{array}{ccccc}
0 & 0 & 0 & 0 & 0 \\ 
0 & \rho _sc_2\kappa _s & 0 & 0 & 0 \\ 
0 & 0 & \sigma _s\widehat{k}_x^2+\eta _s & \sigma _s\widehat{k}_x\widehat{k}%
_y & \sigma _s\widehat{k}_x\widehat{k}_z \\ 
0 & 0 & \sigma _s\widehat{k}_y\widehat{k}_x & \sigma _s\widehat{k}_y^2+\eta
_s & \sigma _s\widehat{k}_y\widehat{k}_z \\ 
0 & 0 & \sigma _s\widehat{k}_z\widehat{k}_x & \sigma _s\widehat{k}_z\widehat{%
k}_y & \sigma _s\widehat{k}_z^2+\eta _s
\end{array}
\right) ,  \label{matD}
\end{equation}
where $\widehat{{\bf k}}=\widehat{{\bf k}}(t)$ is the unit vector along $%
{\bf k}(t)$, $c_1\equiv (\gamma _s-1)\alpha _s^{-1}$, $c_2\equiv
(mn_sC_{v,s})^{-1}$, and $\sigma _s=\frac 13\eta _s+\eta _s^{\prime }$.

To simplify the analysis attention is restricted here and below to spatial
perturbations only along the velocity gradient direction, i.e. ${\bf k=}k\, 
\widehat{{\bf y}}$. In this case the linear hydrodynamic equations have
time-independent coefficients (i.e., ${\bf k}(t)={\bf k)}$, 
\begin{equation}
\partial _t\delta \widetilde{y}_\alpha +F_{\alpha \beta }({\bf k},a)\delta 
\widetilde{y}_\beta =0,\hspace{0.4in}F_{\alpha \beta }({\bf k},a)=\left(
A(a)-ikB(a)+k^2D\right)_{\alpha\beta} .  \label{Feq}
\end{equation}
and the matrices $B(a)$ and $D$ simplify to 
\begin{equation}
B_{\alpha \beta }(a)=\left( 
\begin{array}{ccccc}
0 & 0 & 0 & n_s & 0 \\ 
0 & 0 & -2a\eta _sc_2 & c_1 & 0 \\ 
-\rho _s^{-1}a\eta _{s,n} & -\rho _s^{-1}a\eta _{s,T} & 0 & 0 & 0 \\ 
\rho _s^{-1}p_{s,n} & \rho _s^{-1}p_{s,T} & 0 & 0 & 0 \\ 
0 & 0 & 0 & 0 & 0
\end{array}
\right) ,  \label{matB2}
\end{equation}

\begin{equation}
D_{\alpha \beta }=\left( 
\begin{array}{ccccc}
0 & 0 & 0 & 0 & 0 \\ 
0 & c_2\kappa _s & 0 & 0 & 0 \\ 
0 & 0 & \rho _s^{-1}\eta _s & 0 & 0 \\ 
0 & 0 & 0 & \rho _s^{-1}\left( \frac 43\eta _s+\eta _s^{\prime }\right) & 0
\\ 
0 & 0 & 0 & 0 & \rho _s^{-1}\eta _s
\end{array}
\right) .  \label{matD2}
\end{equation}
Equation (\ref{Feq}) can be solved by Laplace transformation, 
\begin{equation}
\delta \widehat{y}_\alpha ({\bf k},z)=\int_0^\infty dt\,e^{-tz}\delta 
\widetilde{y}_\alpha ({\bf k},t),  \label{lapdef}
\end{equation}
with the result 
\begin{equation}
\delta \widehat{y}_\alpha ({\bf k},z)=[zI+F({\bf k},a)]_{\alpha \beta
}^{-1}\delta \widetilde{y}_\beta ({\bf k},t=0).  \label{soly1}
\end{equation}
The eigenvalues $\omega ^{(i)}({\bf k},a)$ of the matrix $F({\bf k},a)$
define the five simple hydrodynamic poles at $z=-\omega ^{(i)}({\bf k},a)$.
The resulting five exponentials in time represent the hydrodynamic modes for
relaxation of the perturbations around uniform shear flow. At $a=0$
(perturbations of equilibrium) they are the two sound modes, a heat mode,
and a two-fold degenerate shear mode. For finite shear rate, the modes are
more complicated and have qualitative differences. To illustrate, consider
first the case of $k\rightarrow 0$ at fixed, finite $a$, 
\begin{equation}
\omega ^{(i)}({\bf k},a)\rightarrow \left( 
\begin{array}{l}
b_1k^2 \\ 
-\frac 12(1+i\,\sqrt{3})b_2(a)k^{2/3}+\frac 12(1-i\,\sqrt{3}%
)b_3(a)k^{4/3}+b_4k^2 \\ 
-\frac 12(1-i\,\sqrt{3})b_2(a)k^{2/3}+\frac 12(1+i\,\sqrt{3}%
)b_3(a)k^{4/3}+b_4k^2 \\ 
b_2(a)k^{2/3}+b_3(a)k^{4/3}+b_4k^2 \\ 
\left( \eta _s/\rho _s\right) k^2
\end{array}
\right) .  \label{mode2}
\end{equation}
The coefficients in these expressions are real, 
\[
b_1=\left( \eta _{s,T}p_{s,n}-\eta _{s,n}p_{s,T}\right) /m\,p_{s,T},\hspace{0.4in}%
b_2(a)=\left( 2a^2\eta _sp_{s,T}/\rho _s^2C_{v,s}\right) ^{1/3}, 
\]
\[
b_3(a)=\left( 2a^2\eta _s\eta _{s,T}\left( \rho _sC_{v,s}\right)
^{-1}+n_sp_{s,n}+\left( \gamma _s-1\right) \alpha _s^{-1}p_{s,T}\right) /3\rho
_s\,b_2(a), 
\]
\begin{equation}
b_4=\frac 13\left( -\,b_1+\kappa _s\left( \rho _sC_{v,s}\right) ^{-1}+\rho
_s^{-1}\left( 2\eta _s+\sigma _s\right) \right) .  \label{coef}
\end{equation}
There are two diffusive modes, $\sim k^2$, but the other modes are
non-analytic about $k=0$ and represent more complex spatial dependence. This
behavior can be traced to the fact that the matrix $A(a)-ikB(a)$ is not
normal and cannot be diagonalized. Thus, at fixed $a\neq 0$ there is a
crossover in the transformation of $F_{\alpha \beta }({\bf k},a)$ from
normal diagonal form to Jordan form at small $k$. This is reflected in the
eigenvalues if they are expanded in $k$ at fixed $a$, as above. For similar
reasons, care must be used in representing $\delta \widetilde{y}_\alpha (%
{\bf k},t)$ as an expansion in the hydrodynamic modes with constant
coefficients, since at small wavevectors there is a crossover to a mode
expansion whose coefficients have algebraic time dependence.

The two propagating modes in (\ref{mode2}) are unstable, since $b_2(a)>0$.
The above Navier-Stokes analysis applies for small but finite shear rate,
and small $k$ (long wavelength). The expansion in $k$ verifies that the
asymptotic long wavelength modes are always unstable. At larger values of $k$
the modes are again stable, as follows from an exact evaluation of the
eigenvalues. There is a critical wavevector, $k_c(a)$, such that for $%
k>k_c(a)$ the modes are stable whereas they are unstable otherwise. These
qualitative results apply without restriction to the atomic force law,
density, or temperature. Figure \ref{fig_b1} shows $k_c(a)$ as a function of 
$a$ for the special case of hard spheres at three densities, $n^{*}=n\sigma
^3=0.0$, $0.2$, and $0.4$ ($k$ and $a$ measured in units of the inverse
mean free path and mean free time for the hard sphere Boltzmann equation).
The thermodynamic properties are calculated using the Percus-Yevick
approximation, while the transport coefficients are calculated from the
Enskog kinetic theory.

The instability is due to three matrix elements, $B_{42}$, $B_{23}$, and $%
A_{34}$, and is therefore present at order $k$. The density perturbation is
constant to this order and we choose $\delta n=0$ to simplify the
discussion. The relevant variables are then the temperature perturbation $%
\delta T$, the longitudinal velocity perturbation $\delta U_y$, and the
transverse velocity perturbation $\delta U_x$ which to this order obey the
equations 
\begin{equation}
\partial _t\delta T+(\gamma _s-1)\alpha _s^{-1}\partial _y\delta U_y-2a\eta
_s\left( \rho _sC_{v,s}\right) ^{-1}\partial _y\delta U_x=0,
\label{mechanism1}
\end{equation}
\begin{equation}
\partial _t\delta U_x+a\delta U_y-\rho _s^{-1}a\eta _{s,T}\partial _y\delta
T=0,\hspace{0.3in}\partial _t\delta U_y+\rho _s^{-1}p_{s,T}\partial _y\delta T=0.
\label{mechanism}
\end{equation}
The first equation has a coupling to the transverse velocity field due to
the reference shear flow; the second equation provides a feedback to the
temperature equation through the same shear flow. These couplings alone
would  lead simply to a renormalization of the sound velocity. However, the
shear flow also couples the transverse field to the longitudinal field for
an additional feedback mechanism to the temperature equation through the
pressure gradient. This second mechanism is responsible for the instability.
For very long wavelengths the above equations can be simplified to give 
\begin{equation}
\partial _t^3\delta T\sim \left( 2a^2\eta _sp_{s,T}/\rho _s^2C_{v,s}\right)
\partial _y^2\delta T,  \label{mechanism2}
\end{equation}
which exhibits the unstable modes of (\ref{mode2}).

\section{Monte Carlo Simulation}

\label{sec:monte}

The hydrodynamic description can be derived from a more fundamental level of
kinetic theory. In principle, this also allows derivation of hydrodynamic
equations without the restriction of the Navier-Stokes approximation to
small shear rates. A model kinetic theory for the practical calculation of
such generalized hydrodynamic equations is given in Ref.\ \cite{mirim:3}.
The resulting equations are limited to long wavelengths, as in the
Navier-Stokes case, but the reference state of uniform shear flow can have a
large shear rate. An analysis of the hydrodynamic modes shows that there is
a critical wavevector similar to that of Figure \ref{fig_b1}. It is possible
to test this hydrodynamic description by a direct simulation of the more
fundamental solution to the kinetic equation. For practical reasons the
simulation is more efficient at larger wavevectors and shear rates than can
be justified by Navier-Stokes hydrodynamics, and this is the primary reason
for considering the more complex generalized hydrodynamics.

The kinetic equation is a single relaxation time BGK equation \cite{BGK}
given by 
\begin{equation}
\left( \frac \partial {\partial t}+{\bf v}\cdot {\bf \nabla }_r+ {\bf \nabla 
}_vm^{-1}\cdot {\bf F}_{\text{ext}}\right) f({\bf r},{\bf v},t)=-\nu (f({\bf %
r},{\bf v},t)-f_\ell ({\bf r},{\bf v},t)).  \label{BGKeq}
\end{equation}
Here ${\bf F}_{\text{ext}}$ is the external force representing the
thermostat, $f_\ell $ is the local equilibrium distribution, and $\nu $ is
an average collision rate. The exact stationary solution for uniform shear
flow in the presence of a thermostat has been studied in detail \cite{usf}.
A variant of the Chapman-Enskog method can be used to study normal solutions
for states near uniform shear flow, and to obtain the corresponding
hydrodynamic equations for small perturbations relative to this state. The
complete details of this solution and the hydrodynamic modes as a function
of the shear rate can be found in \cite{mirim:3}. The method of simulation
is a variant of the Bird Direct Simulation method \cite{bird}. This is a
Monte Carlo technique composed of two parts at each discrete time step: free
streaming and collision. The volume of the system is divided into cells of
dimension smaller than the mean free path, and $N$ particles are introduced
at $t=0$ with positions and velocities sampled statistically from a
specified initial distribution function. The distribution of particles is
calculated at $t=\Delta t $, with $\Delta t $ much smaller than the mean
free time, as follows. First the particles are displaced by $\{\Delta {\bf r}%
_\mu ={\bf v} _\mu \Delta t \}$. Next the velocity of each particle $\mu $
is replaced with probability $\nu (n_\mu ,T_\mu)\Delta t $ by a random
velocity sampled from the local equilibrium distribution, $%
f_\ell(\{n_\mu,T_\mu,{\bf U}_\mu\};{\bf v})$. Here $n_\mu$, $T_\mu$, and $%
{\bf U}_\mu$ are the density, temperature, and flow velocity in the cell
containing particle $\mu$. In this collision stage, strict conservation of
momentum and energy may be violated due to statistical fluctuations. To
compensate for this artificial effect, the velocities of the particles in
each cell are conveniently displaced and rescaled. The whole process is then
repeated for each subsequent time step. In this way, the one particle
distribution function $f({\bf r},{\bf v},t)$ (coarse grained over the cells)
is determined, from which the hydrodynamic fields can be computed directly
as averages.

In our simulations we have considered a system of size $L=2\pi /k$, with $%
k=0.1(v_0t_0)^{-1}$, along the $y$-direction at a shear rate $a=0.5t_0^{-1}$%
. Here $t_0=1/\nu (n_s)$ is the mean free time with $\nu \propto n$ (Maxwell
molecules) and $v_0=(2k_BT_s/m)^{1/2}$ is the thermal velocity. In the
remainder of this section, we take $t_0=1$, $v_0=1$, $T_s=1$, and $n_s=1$.
Lees-Edwards boundary conditions are used to drive the shear flow \cite
{edwards}. These are simple periodic boundary conditions on both the
position and velocity variables in the local Lagrangian frame at $y=\pm L/2$%
. Both local and global thermostats have been studied. The local thermostat
is the same as that of section \ref{sec:navier} except
that the pressure tensor is
no longer limited to its Navier-Stokes form. In addition, two global
thermostats are considered for more efficient implementation of the
simulation. These are described in the Appendix. The analysis there and the
results of the simulation show that the instability is not sensitive to the
choice of thermostat. Consequently, only the results using the global
thermostats are presented here. Starting from a distribution corresponding
to uniform shear flow \cite{MSG96}, the initial condition has been prepared
by displacing and rescaling velocities so that 
\begin{equation}
\delta U_{x,y}(y,0)=-\delta \widetilde{U}_{x,y}(0)\sin ky,\quad \delta
U_z(y,0)=\delta n(y,0)=\delta T(y,0)=0,  \label{d1}
\end{equation}
with $\delta \widetilde{U}_y(0)=0.1$ and $\delta \widetilde{U}_x(0)=-0.03$.
The technical parameters of the simulations are $N=628\,000$ particles, a
time step $\Delta t=0.02$, and a cell width $\Delta L=0.05$. The data have
been averaged over $10$ different realizations in the simulations of Figs.\ 
\ref{fig_d1}, \ref{fig_d3}, \ref{fig_d6}, and \ref{fig_d5}, and over 2 realizations in
those of Figs.\ \ref{fig_d2} and \ref{fig_d4}.

First we consider the short time dependence at a fixed position $y=-L/4$.
Figure \ref{fig_d1} shows $\delta \widetilde{U}_x(t)=\delta U_x(-L/4,t)$ and
$\delta \widetilde{U}_y(t)=\delta U_y(-L/4,t)$.
The dashed lines are the results from the hydrodynamic analysis of the BGK
model near uniform shear flow \cite{mirim:3}. The good agreement up to $%
t\approx 50$ shows that the instability is not just a consequence of the
assumptions behind the hydrodynamic description. The subsequent differences
between simulation and theory are due to the fact that the latter is limited
to small deviations from uniform shear flow. For longer times we find both
large amplitudes for $\delta y_\alpha $ and large deviations of the
distribution function from that of the unperturbed state.

To investigate the asymptotic state the system, we have performed the
simulations for much longer times. Figure \ref{fig_d2} shows $\delta 
\widetilde{U}_x(t)$ and $\widetilde{T}(t)=T(-L/4,t)$ for
time up to $t=2000$. Both the velocity and the temperature
oscillate in time and
are modulated by a slowly varying amplitude relative to their
asymptotic average values. The maximum value $\widetilde{T}\approx 9$
corresponds to $k\approx 0.3$, which is close to the value $k_c(a)$ at which
the uniform shear flow with $a=0.5$ would become marginally stable \cite
{mirim:2}. Thus the initial dynamics tends toward stabilization but does not
ever cross over into the stable domain. Consequently, the asymptotic
dynamics is not simply that of stationary uniform shear flow at a different
temperature, but rather a quasi-stationary state with different spatial
structure.

The thermostat used in the above simulations allows a global change in the
average temperature, which is responsible for the amplitude modulations at
long times. In order to have a more controlled asymptotic state, we have
considered a variation of this thermostat that maintains the average
temperature constant (see the Appendix). The quantities $\delta \widetilde{U}%
_y(t)$ and $\widetilde{n}(t)=n(-L/4,t)$ are plotted in Fig.\ \ref{fig_d3}.
After a transient period of length $t\approx 100$, a stable oscillatory
behavior of the velocity appears with a period $\tau \simeq 54$. The shape
of $\delta \widetilde{y}_\alpha $ over one cycle is shown in Fig.\ \ref
{fig_d4}. Here $t^{\prime }=t-t_{\text{tr}}$, where $t_{\text{tr}}=463.4$,
and the curves are averages over 20 successive cycles. The value of $t_{%
\text{tr}}$ has been chosen as to assure that the transient time is over and
also with the criterion that $\delta \widetilde{U}_x=0$ at $t^{\prime }=0$.
Inspection of the results shows several regularities. First, the following
symmetry relation appears: 
\begin{equation}
\delta \widetilde{y}_\alpha (t^{\prime }+\tau /2)=\pm \delta \widetilde{y}%
_\alpha (t^{\prime }),  \label{d2}
\end{equation}
where the minus sign applies to the velocity and the plus sign applies to
the density and the temperature. Next, at times $t^{\prime }\simeq 0.36\tau
,0.86\tau $, where $\delta \widetilde{U}_x$ have extrema, $\delta \widetilde{%
U}_y$, $\delta \widetilde{T}$, and $\delta \widetilde{n}$ seem to have
nodes. Note also that the vector $\delta \widetilde{{\bf {U}}}$ rotates
anti-clockwise and that most of the time $\delta \widetilde{T}>0$ is
correlated to $\delta \widetilde{n}>0$ and $\delta \widetilde{T}<0$ is
correlated to $\delta \widetilde{n}<0$.

The above results indicate the wave character of the asymptotic state. To
confirm this point, we have analyzed the profiles of the hydrodynamic
quantities at relevant times. The results are consistent with two
independent invariance relations: 
\begin{equation}
\delta y_\alpha (y,t^{\prime })=\delta y_\alpha (y+L/2,t^{\prime }+\tau /2),
\label{d3}
\end{equation}
\begin{equation}
\delta y_\alpha (y,t^{\prime })=\pm \delta y_\alpha (-y,t^{\prime }).
\label{d4}
\end{equation}
Their combination yields $\delta y_\alpha (y,t^{\prime })=\pm \delta
y_\alpha (-y-L/2,t^{\prime }+\tau /2)$, which implies Eq.\ (\ref{d2}).
Figure  \ref{fig_d6} shows the spatial variation of $\delta
n(y,t^{\prime })$ and $\delta T(y,t^{\prime })$ at $t^{\prime }=0,0.14\tau
,0.25\tau ,0.36\tau ,\text{ and }0.5\tau $. Not shown are times $0.5\tau
<t^{\prime }<\tau $ because they can be reproduced by use of the relation (%
\ref{d3}).
As observed in Fig.\ \ref{fig_d4} for the special
point $y=-L/4$ it is seen that a high (low) temperature is generally
correlated to a high (low) density. The spatial distribution of the
temperature is highly non-uniform even though the thermostat maintains a 
constant average temperature. Figure \ref{fig_d5} shows a vector
representation of the components of $\delta {\bf U}$ throughout the system
at the values of $t^{\prime }$. As anticipated from Fig.\ \ref{fig_d4}, $%
\delta {\bf U}$ rotates anti-clockwise throughout the system.
The layers $y=0,\pm L/2$ are
always nodes of $\delta {\bf {U}}$ and extrema of $%
\delta T$ and $\delta n$.
The pattern indicated by Figs.\ \ref{fig_d6} and \ref{fig_d5} can be
described as a periodic  standing wave represented by the superposition of
two symmetrical waves travelling in opposite directions.

In summary, for initial values of $k$ and $a$ in the predicted unstable
domain small perturbations of the hydrodynamic fields grow according to the
linear hydrodynamic equations for $t<50$. Subsequently, non-linear
effects invalidate this theoretical analysis. The simulations show a
transient period up to about $t\approx 100$, after which a quasi-stationary
state is observed for $100<t\leq 2000$. In this asymptotic state the vector
quantities oscillate at a period approximately twice that of the scalar
fields. The oscillations are spatially non-uniform for all fields considered.
\section{Molecular Dynamics Simulation}

\label{sec:mole}

The most extensive prior studies of uniform shear flow have been for dense
systems via molecular dynamics (MD) simulations. No signature of the
instability discussed here has been noted in these previous results. There
are two significant differences between the theory and simulation discussed
above and the detailed implementation of earlier MD simulations. The first,
and perhaps most important, has been the consideration of small system sizes
relative to the wavelengths necessary to see the instability. Typically,
system size is determined by the simulation time such that a sound wave will
not traverse the system and generate correlations. At high densities this
has led to consideration of system sizes small compared to the critical
wavevector for instability. A second difference from the discussion above is
the method for imposing a thermostat. In MD simulations it is efficient to
impose the temperature control by a global re-scaling of the velocities only
after as many as $100$ collisions. However, at the shear rates considered
this implies significant heating between applications of the thermstat and
the temperature is more like a sawtooth in time rather than constant. Thus
it is difficult to give a direct theoretical correspondence to the MD
simulation results.

To demonstrate the existence of the instability using MD simulations we have
considered a rectangular unit cell with one side expanded $15$ times larger
than the other two ( $1620$ hard spheres of diameter $\sigma $ at a density
of $n\sigma ^3=0.5$ in a cell of size $6\sigma \times 90\sigma \times
6\sigma )$. The Lees-Edwards boundary conditions are applied on the $x,z$
surfaces so that the velocity gradient is along the larger dimension of the
system, allowing study of much longer wavelengths in the direction of the
gradient than have previously been considered. A second difference from
previous simulations is a more frequent application of the velocity scaling
to control the themperature and better represent continuous cooling. In our
simulations the velocities are rescaled whenever the temperature differs by
5\% from its set value. The simulation cell is divided into a fixed number
of subcells, the local velocity field in each subcell is calculated and the
excess kinetic energy computed relative to the local velocity field. The
velocities of the particles in each subcell are then rescaled so that the
total excess kinetic energy of each subcell is equal to the set value.
The amount
of rescaling is determined locally, and thus corresponds to a local
thermostat, $\lambda \left( {\bf r},t\right) =\lambda \left[ n\left( {\bf r}%
,t\right) ,T\left( {\bf r},t\right) \right] $, as discussed in the Appendix.
Our implementation follows Hess \cite{shear} in that we assume
uniformity of all quantities in the directions perpendicular to the velocity
gradient so that the subcells are thin slabs and the only spatial variations
are in the direction of the velocity gradient, as in the Monte Carlo
simulations. Due to the large number of particles in the simulations, the length
of the simulations is relatively modest $t\approx 5\times 10^5$ collisions. The
shear rate was fixed at $a=1.77\sqrt{k_BT/m\sigma^2}$
and an initial perturbation with $k_y=
2\pi/L_y$ was monitored. Our theoretical estimates indicate
this should correspond to conditions of instability. Figure \ref{fig_jfl_1}\
shows the $x$-component of the velocity field growing steadily throughout the
simulation, clearly indicating the instability. Conversely, perturbations
with a wavelength one quarter of this value appear to be stable as expected.
Similar results are observed for the density field as well. Our primary
conclusion from these preliminary results is that the instability can be
observed and studied by MD simulations if larger system sizes are considered
and more care is taken with application of the thermostat.

\section{Discussion}

\label{sec:dis}

Uniform shear flow has been a prototype state for the study of fluids far
from equilibrium, using both theoretical and computer simulation methods.
Until recently, it has been assumed that this state is stable except at high
densities and short wavelengths. The new results reported here and in \cite
{mirim:2,mirim:3} show that this simple macroscopic state is unstable at
sufficiently large wavelengths. Previous studies via simulation have not
seen this instability due to finite system sizes. However, theoretical
analysis at the hydrodynamic level clearly shows the mechanism and parameter
space for this instability. In the present work this theoretical analysis is
tested both qualitatively and quantitatively. At the qualitative level, both
Monte Carlo simulations of a kinetic theory description and molecular
dynamics simulation of the Newtonian dynamics show clearly that this flow is
unstable at long wavelengths. The Monte Carlo simulation also confirms
quantitatively the predictions of the theory on the time scale for which the
linear analysis is valid. At longer times the Monte Carlo simulation shows
clearly that the asymptotic state is spatially non-uniform with a periodic
variation in time. A precise theoretical description of this final state has
not been developed at this time.

The analysis here has been limited to spatial perturbations along the
gradient of the velocity in the stationary state. More general perturbations
will lead to more complex flows due to the coupling to the convective flow.
A theoretical analysis of this more general case is in progress but is
significantly more complex. Due to the fact that the $k=0$ matrix of the
hydrodynamic equations is non-diagonalizable (non-normal), the corresponding
eigenvalues do not fix a unique eigenspace. The resulting dynamics is not
simply a superposition of decaying or growing modes, but rather includes as
well algebraic growth. The resulting analysis of conditions for instability
is more complex and will be reported elsewhere.

\acknowledgments

The research of M.L. and J.D. was supported in part by NSF grant PHY
9312723. The research of J.M.M. and A.S. was supported in part by the DGICYT
(Spain) and by the Junta de Extremadura-Fondo Social Europeo through Grants
No. PB94-1021 and No. EIA94-39, respectively. Partial support for this
research also was provided by the Division of Sponsored Research at the
University of Florida.
\appendix
\section*{Role of the Thermostat}

\label{appendix:A}

The analysis of section \ref{sec:navier} made use of a specific choice for
the thermostat. Other choices are possible and more convenient for computer
simulation. In this work we use two different types of thermostats, both
obtained from an external force at the microscopic level of the form 
\begin{equation}
{\bf F}_{\text{ext}}({\bf r},t)=-m\lambda (n({\bf r},t),T({\bf r},t))[{\bf v}%
-{\bf U}({\bf r},t)].  \label{force}
\end{equation}
The corresponding source term $w$ in the temperature equation (\ref{2}) is 
\begin{equation}
w(n({\bf r},t),T({\bf r},t))=-2K(n({\bf r},t),T({\bf r},t))\lambda (n({\bf r}%
,t),T({\bf r},t)),  \label{source}
\end{equation}
where $K(n({\bf r},t),T({\bf r},t))$ is the kinetic energy density.
% and the
%second line shows terms through linear order in the deviations from uniform
%shear flow.
The thermostat parameter, $\lambda (n({\bf r},t),T({\bf r},t))$,
is always chosen to ensure constant temperature in the reference state, $%
\lambda _s=\lambda (n_s,T_s)=-aP_{s,xy}/2K_s$, as follows from (%
\ref{5}). In this work we use three different thermostats, one local and two
global. The local thermostat is defined by $\lambda (n({\bf r},t),T({\bf r}%
,t))\rightarrow \lambda _s+\lambda _{s,n}\delta n({\bf r},t)+\lambda
_{s,T}\delta T({\bf r},t)$, to linear order, so that 
\begin{equation}
w_1(n({\bf r},t),T({\bf r},t))\rightarrow -2K_s\lambda _s-2\left( K\lambda
\right) _{s,n}\delta n({\bf r},t)-2\left(K\lambda\right) _{s,T}\delta T(%
{\bf r},t).  \label{source1}
\end{equation}
The coefficients $\lambda _{s,n}$ and $\lambda _{s,T}$ are chosen such that
all viscous heating proportional to $\delta n({\bf r},t)$ and $\delta T({\bf %
r},t)$ is compensated by the source term. While there is still a local
temperature variation, this thermostat holds the average temperature
constant. It is this local thermostat that is used in section \ref
{sec:navier}, with $P_{xy}$ approximated by its Navier-Stokes form. The
other two thermostats considered are global, i.e. the parameter $\lambda $
is spatially constant. The simplest case is $\lambda =\lambda _s$ with a corresponding
source term 
\begin{equation}
w_2(n({\bf r},t),T({\bf r},t))\rightarrow -2K_s\lambda _s
-2\lambda _s K_{s,n}\delta n({\bf r},t)-2\lambda _sK_{s,T}\delta T({\bf r}%
,t).   \label{source2}
\end{equation}
In this case there are greater effects due to viscous heating by the
perturbations than with the choice $w_1$, and even the average temperature
of the system changes except at the stationary state. A second global
thermostat eliminates this global change using the form, $\lambda
\rightarrow \lambda _s+\lambda _{s,n}\delta \overline{n}+\lambda
_{s,T}\delta \overline{T}$, where $\delta \overline{n}$ and $\delta 
\overline{T}$ are the volume averaged density and temperature. The source
term in this case is 
\begin{equation}
w_3(n({\bf r},t),T({\bf r},t))\rightarrow w_2(n({\bf r},t),T({\bf r}%
,t))-2K_s\left[ \lambda _{s,n}\delta \overline{n}+\lambda _{s,T}\delta 
\overline{T}\right].   \label{source3}
\end{equation}
In Fourier representation the sources $w_2$ and $w_3$ are the same except
for the $k=0$ dynamics.

Both theory and the Monte Carlo simulations of section \ref{sec:monte}
indicate that the details of the perturbed dynamics depend on the thermostat
used, but that the mechanisim for the instability is insensitive to the
choice of thermostat. To illustrate this we repeat the analysis of section 
\ref{sec:navier} using the global thermostat (\ref{source2}) (again in the
Navier-Stokes limit). There are two new terms in the temperature equation.
One is due to temperature perturbations and leads to viscous heating even
for uniform perturbations. The second is due to density perturbations and
provides a direct $k$ - independent coupling of the temperature and density
equations (at zero shear rate the density and temperature are only coupled
indirectly via the gradients of the longitudinal velocity component and the
pressure). The matrices $B(a)$ and $D$ are unchanged from section \ref
{sec:navier}, but these new couplings replace $A_{21}$ and $A_{22}$ in (\ref
{matA}) with non-zero values proportional to $a^2$, 
\begin{equation}
\left( 
\begin{array}{c}
A_{21}(a) \\ 
A_{22}(a)
\end{array}
\right) \rightarrow \left( 
\begin{array}{c}
-c_2\left( a^2\eta _{s,n}+w_n\right) \\ 
-c_2\left( a^2\eta _{s,T}+w_T\right)
\end{array}
\right) ,
\end{equation}
where $w_n\equiv \partial w(n_s,T_s)/\partial n_s$, $w_T\equiv \partial
w(n_s,T_s)/\partial T_s$. The matrix $A(a)-ikB(a)$ is now diagonalizable and
an expansion of the hydrodynamic modes in powers of $k$ rather than $k^{2/3}$
is obtained, 
\begin{equation}
\omega ^{(i)}({\bf k},a)\rightarrow \left( 
\begin{array}{l}
A_{22}(a)+d_1(a)k^2 \\ 
d_2(a)k^2 \\ 
ic(a)k-\left( d_3(a)/a^2-d_4(a)\right) k^2 \\ 
-ic(a)k-\left( d_3(a)/a^2-d_4(a)\right) k^2 \\ 
\left( \eta _s/\rho _s\right) k^2
\end{array}
\right) ,  \label{mode3}
\end{equation}
with the coefficients, 
\[
d_1(a)=\kappa _sc_2+\frac{2a^2c_2\eta _s\eta _{s,T}+c_1p_{s,T}}{\rho _sc_2(\eta
_{s,T}a^2+w_T)}+\frac{p_{s,T}(n_s\eta _{s,n}a^2+n_sw_n+2\eta _sa^2)}{\rho
_sc_2(\eta _{s,T}a^2+w_T)^2}, 
\]
\[
d_2(a)=-\frac{\eta _s\left[ p_{s,n}\left( a^2\eta _{s,T}-w_T\right) +p_{s,T}\left(
-a^2\eta _{s,n}+w_n\right) \right] }{m\left[ n_sp_{s,n}\left( a^2\eta
_{s,T}+w_T\right) -p_{s,T}\left( a^2n_s\eta _{s,n}+n_sw_n+2\eta _sa^2\right)
\right] }, 
\]
\[
c^2(a)=\frac{\left[ n_sp_{s,n}\left( a^2\eta _{s,T}+w_T\right) -p_{s,T}\left(
a^2n_s\eta _{s,n}+n_sw_n+2\eta _sa^2\right) \right] }{\rho _s\left( a^2\eta
_{s,T}+w_T\right) }, 
\]
\[
d_3(a)=\frac{p_{s,T}a^2}{2\rho _sc_2\left( a^2\eta _{s,T}+w_T\right) }\left[ c_1+%
\frac{\left( a^2n_s\eta _{s,n}+n_sw_n+2\eta _sa^2\right) }{\left( a^2\eta
_{s,T}+w_T\right) }\right] , 
\]
\begin{equation}
d_4(a)=-\frac 12\,d_2(a)-\frac{a^2\eta _s\eta _{s,T}}{\rho _s\left( a^2\eta
_{s,T}+w_T\right) }+\frac{2\eta _s+\sigma _s}{2\rho _s}.
\end{equation}

There are two diffusive modes, two propagating sound modes, and a
time-modulated diffusive mode. The sound modes are unstable at long
wavelengths for shear rates satisfying $d_3(a)-a^2d_4(a)>0$. This
possibility has been explored in reference \cite{mirim:2} for the special
case of hard spheres. In that case both $d_3(a)$ and $d_4(a)$ are positive
and the modes are unstable for sufficiently small shear rates. The
corresponding critical wavevector, $k_c(a)$, for this thermostat is
determined from the exact eigenvalues and shown in Figure \ref{fig_c1} for
the same densities as in Figure \ref{fig_b1}. The domain of instability at
long wavelengths is similar to that of Figure \ref{fig_b1} using the
thermostat $w_1$.

The thermostat of this Appendix allows greater viscous heating than that of
section \ref{sec:navier} for states perturbed from uniform shear flow. The
hydrodynamic modes are quite different, reflecting a sensitivity to the
choice of thermostat. These qualitative differences can be traced to the
mathematical differences between diagonalizable and non-diagonalizable
matrices, $A_{\alpha \beta }$. Nevertheless, the mechanism for the
instability described at the end of section \ref{sec:navier} remains
effective in both cases. These conclusions are not limited to the
Navier-Stokes approximation, but are confirmed as well for the kinetic model
results described in section \ref{sec:monte} for larger shear rates.

\nopagebreak

\begin{figure}[tbp]
\caption{Critical lines for stability for hard spheres at $n^*=0.0$
(solid curve), $n^*=0.2$ (dashed curve), and $n^*=0.4$ (dashed-dotted curve)
with the local thermostat.}
\label{fig_b1}
\end{figure}
\begin{figure}[tbp]
\caption{ Plot of  $\delta \widetilde{U}_x(t)$ and 
$\delta \widetilde{U}_y(t)$. The solid
lines are simulation results and the dashed lines correspond to the analysis
of Ref.\ \protect\cite{mirim:3}.}
\label{fig_d1}
\end{figure}
\begin{figure}[tbp]
\caption{ Plot of $\delta \widetilde{U}_x(t)$ and $\delta \widetilde{T}(t)$.}
\label{fig_d2}
\end{figure}
\begin{figure}[tbp]
\caption{ Plot of $\delta \widetilde{U}_y(t)$ and $\delta \widetilde{n}(t)$.}
\label{fig_d3}
\end{figure}
\begin{figure}[tbp]
\caption{ Plot of $\delta \widetilde{y}_\alpha (t^{\prime })$ over one cycle.}
\label{fig_d4}
\end{figure}
\begin{figure}[tbp]
\caption{ $T(y,t^{\prime })$ and $n(y,t^{\prime })$ as a function of $y$ for
(from top to bottom)  $t^{\prime }=0,0.14\tau ,0.25\tau ,0.36\tau ,\text{
and }0.5\tau $.}
\label{fig_d6}
\end{figure}
\begin{figure}[tbp]
\caption{ Vector plot representing $\delta {U}_x(y,t^{\prime })$ and $\delta 
{U}_y(y,t^{\prime })$ at  $t^{\prime }=0,0.14\tau
,0.25\tau ,0.36\tau ,\text{ and }0.5\tau $.}
\label{fig_d5}
\end{figure}
\begin{figure}[tbp]
\caption{ The long-wavelength component of the velocity in the flow
direction as a function of time from molecular dynamics simulation.
Units are $m=\sigma=k_BT=1$.}
\label{fig_jfl_1}
\end{figure}
\begin{figure}[tbp]
\caption{Critical lines for stability for hard spheres for the same
densities as in Fig.\ \protect\ref{fig_b1}
with the global thermostat (\protect\ref{source2}).}
\label{fig_c1}
\end{figure}

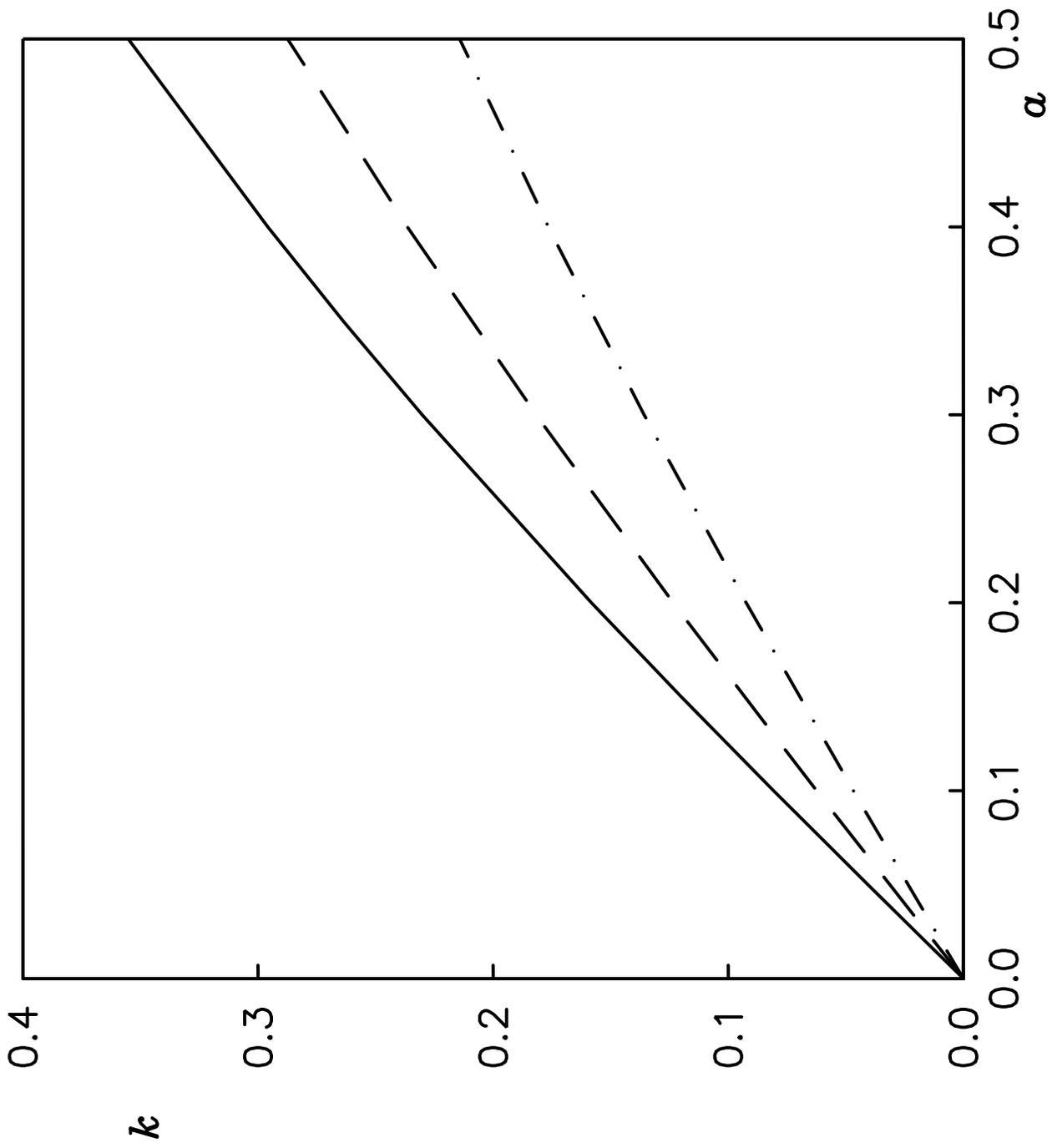
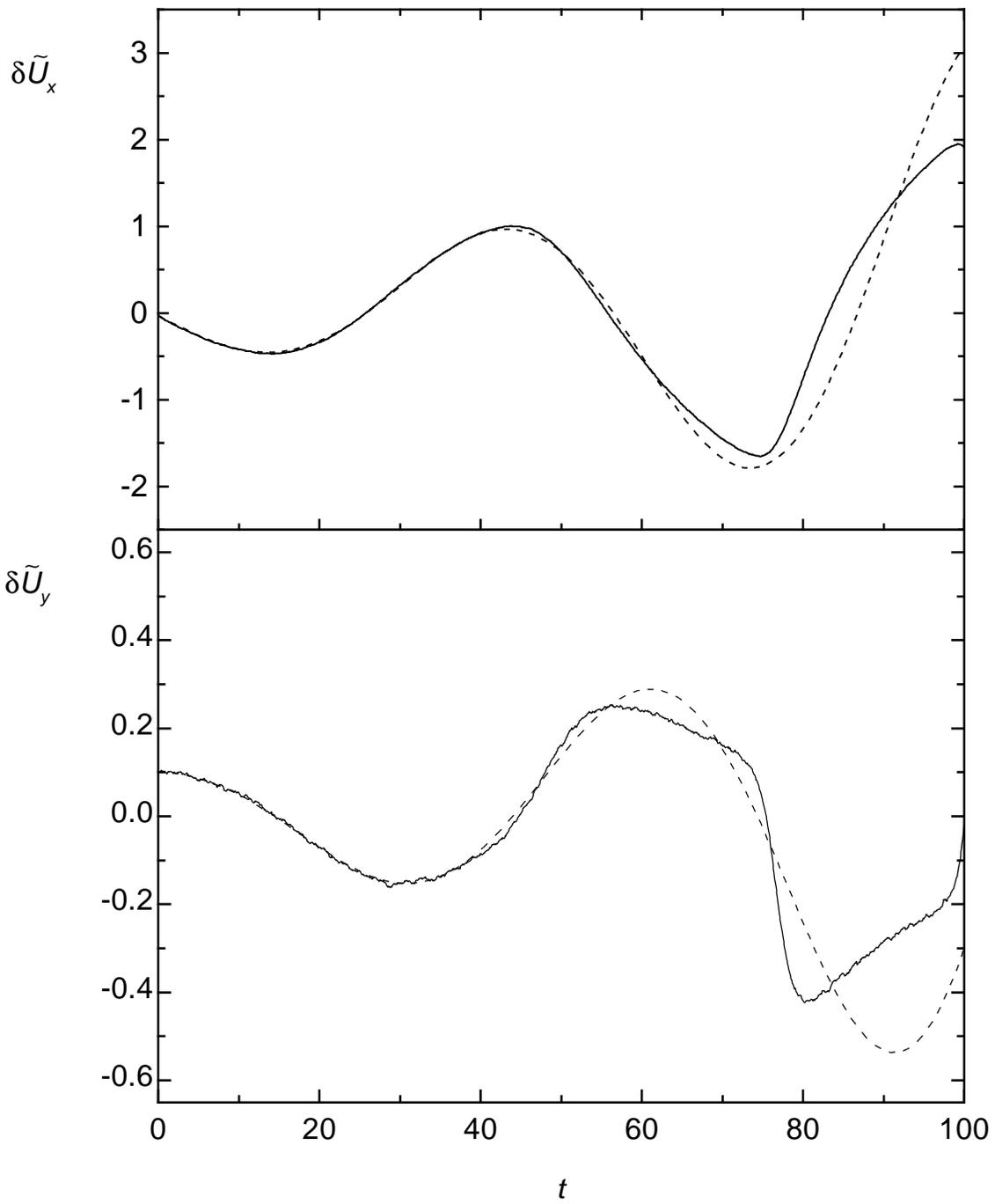
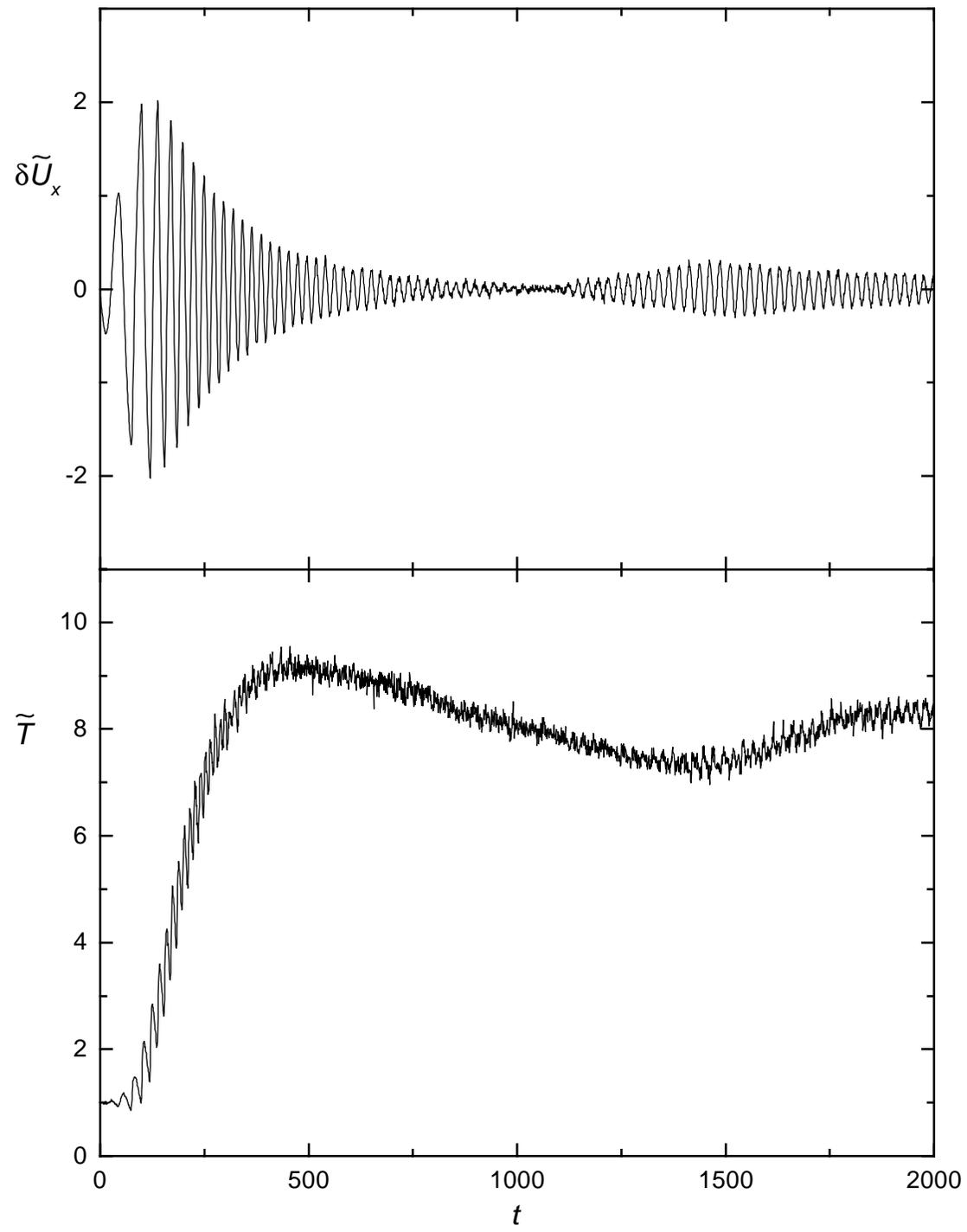
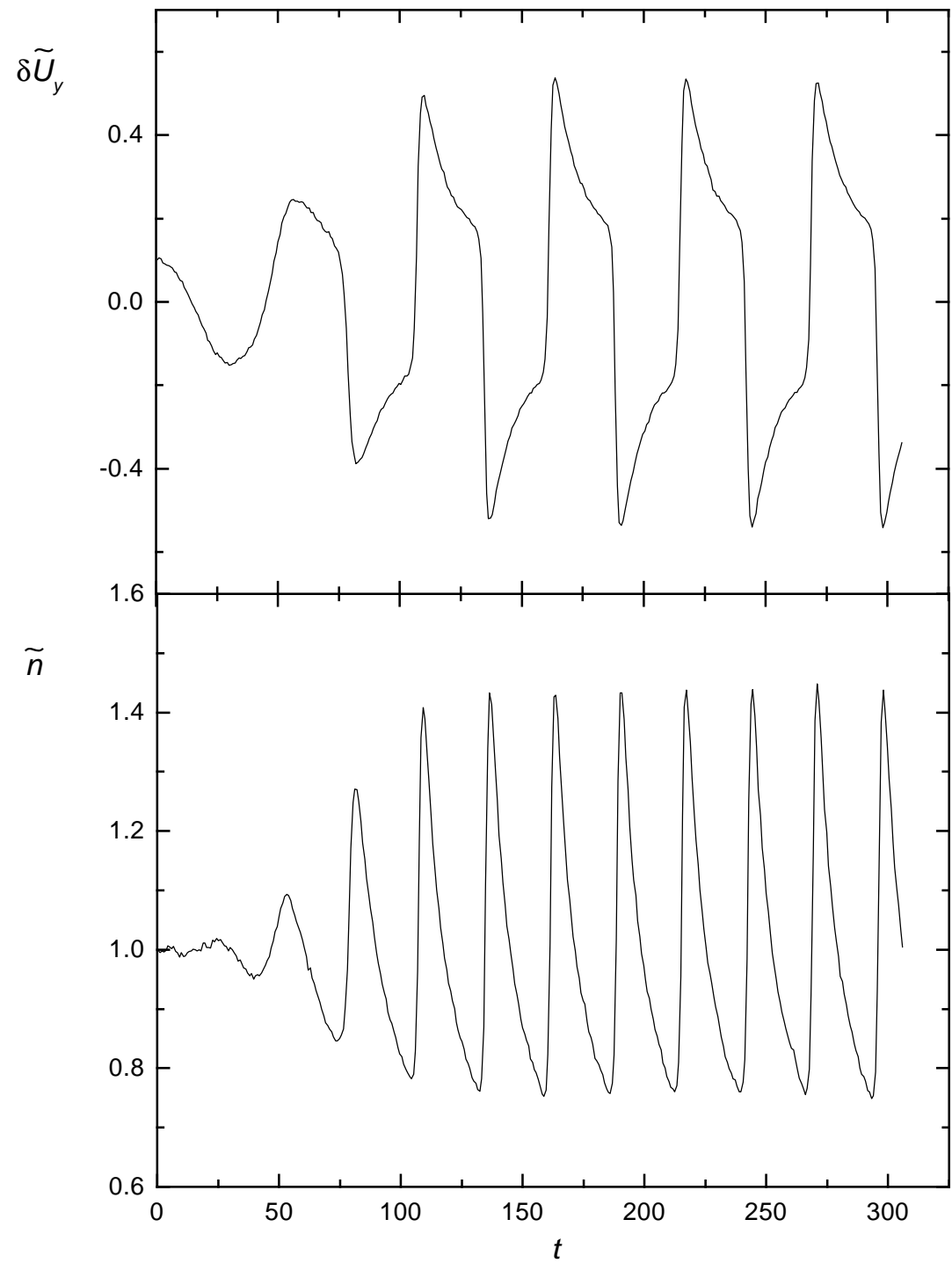
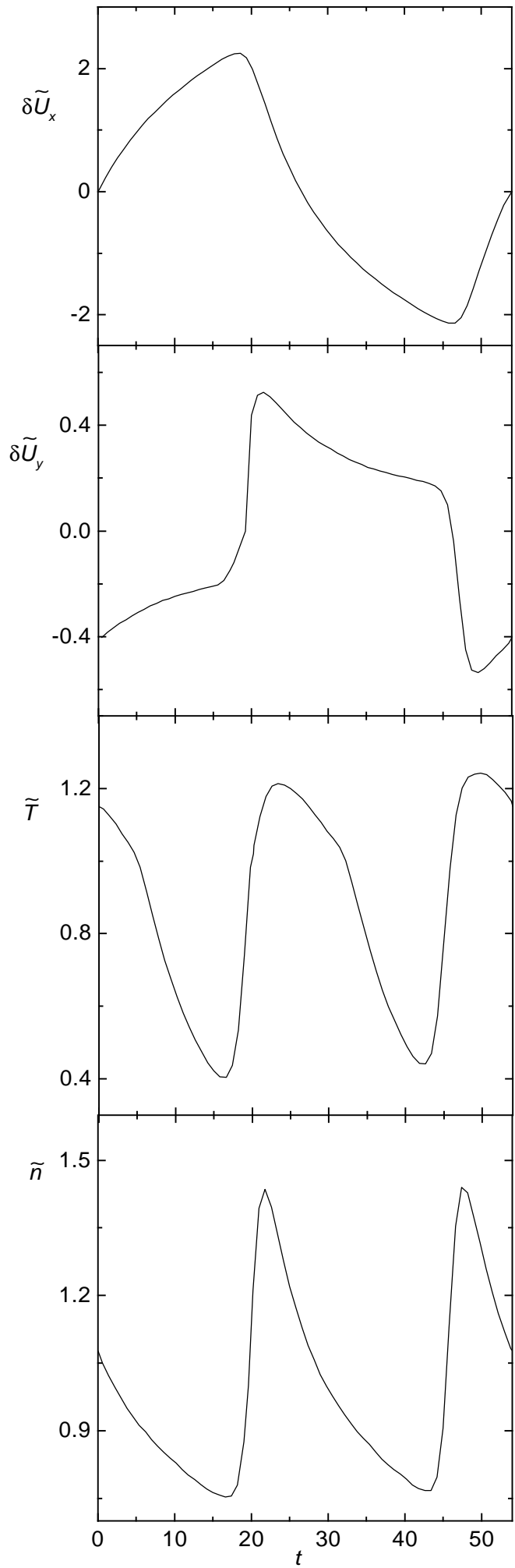
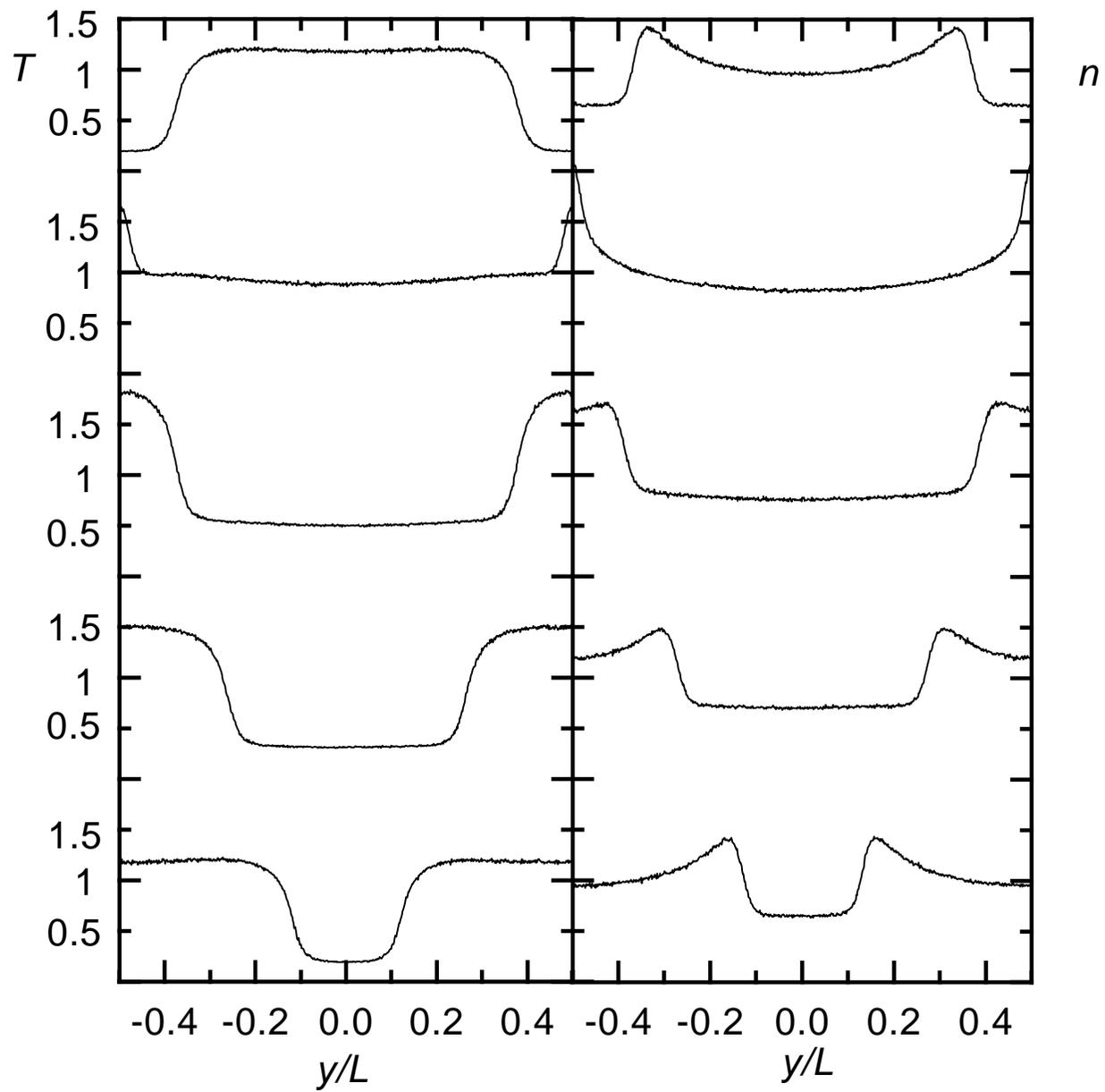
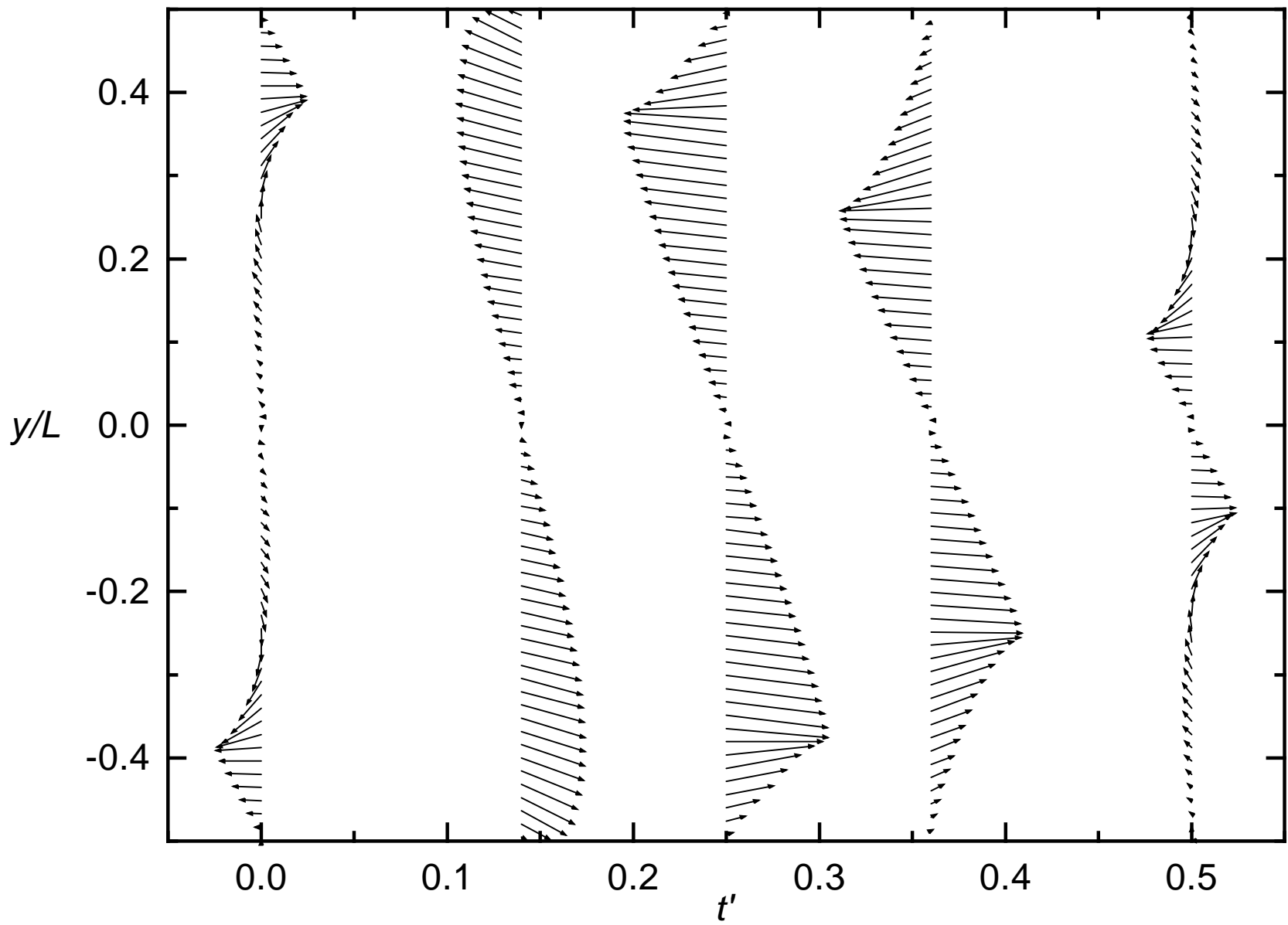
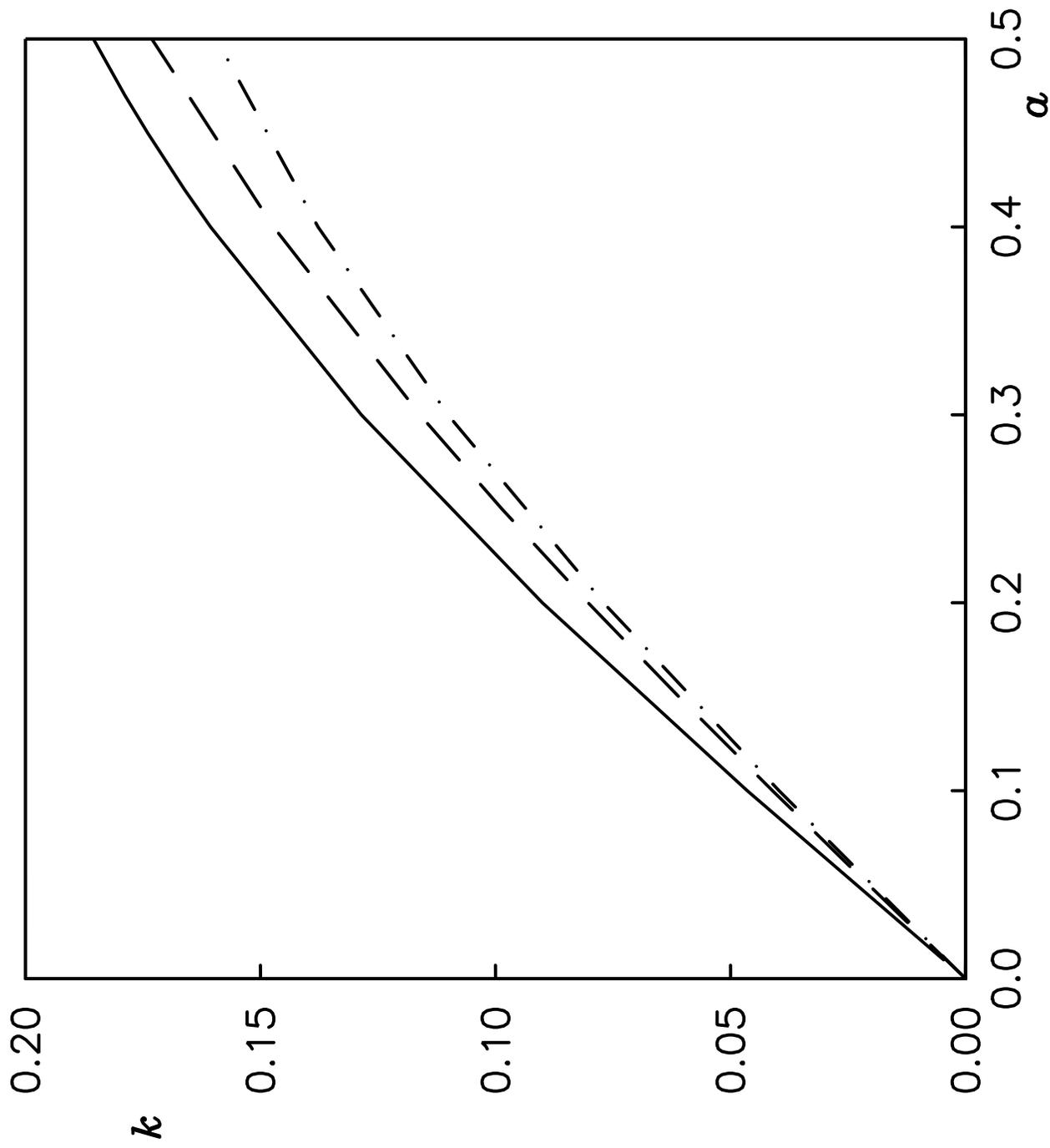
\end{document}